\documentclass{aastex631}

\usepackage{bm}
\usepackage[T1]{fontenc}
\usepackage{booktabs}
\usepackage{threeparttable}
\usepackage{multirow}
\usepackage{amsmath}
\usepackage{makecell}
\usepackage{numprint}
\usepackage{tabularx}
\usepackage{xcolor}
\usepackage{caption}

\begin{document}

\title{Discrepancies in Pedestrian Crossing of Static vs. Dynamic Crowds: An Experimental Study}
 
\author{Jinghui Wang}

\affiliation{School of Safety Science and Emergency Management\\
Wuhan University of Technology\\
Wuhan, China}

\author{Yajuan Jiang}

\affiliation{School of Safety Science and Emergency Management\\
Wuhan University of Technology\\
Wuhan, China}

\author{Xiaoying Zhang}

\affiliation{Institute of Information Technology\\
Hainan Vocational University of Science and Technology\\
Haikou, China}

\author{Fangwei Deng}

\affiliation{School of Safety Science and Emergency Management\\
Wuhan University of Technology\\
Wuhan, China}

\author{Wei Lv}
\altaffiliation{{\url{weil@whut.edu.cn}}}
\affiliation{School of Safety Science and Emergency Management\\
Wuhan University of Technology\\
Wuhan, China}

\begin{abstract}
In this paper, we investigate disparities in pedestrian crossing behaviors within static and dynamic crowds through experimental analysis. First, qualitative trajectory observations revealed significant behavioral differences in static and dynamic contexts. To quantitatively assess these discrepancies, we introduced a density metric termed the swarm factor. In static contexts, limited variations in speed and swarm factor were observed, which may contribute to the formation of cross-channels, a phenomenon of pedestrian self-organization (tactical level). In contrast, speed and swarm factor exhibited inverse synchronization in dynamic contexts, indicating density-dependent behavioral adaptation (operational level). Finally, orthogonal velocity analysis demonstrated a fundamental pattern in crossing motions: as global density increased, both instantaneous velocity and crossing velocity decreased, while transverse velocity remained stable.

\end{abstract}


\section{Introduction} 
\label{section1}

With the frequent occurrence of crowd crush accidents significantly impacting public safety, research on crowd behavior has become crucially important \citep{feliciani2023trends}. The complicated mechanism of the crowd has caused huge challenges to the relevant investigation. Common forms of cross motion include unidirectional flow \citep{ma2015simulation}, bidirectional flow \citep{feliciani2016empirical}, cross-flow \citep{cao2017fundamental,zanlungo2023macroscopica}, and bottleneck flow \citep{seyfried2009new}, among others. Researchers have conducted classified studies for the investigation of  motion patterns, involving the stop-and-go motion \citep{portz2011analyzing}, lane formation \citep{feliciani2016empirical}, arching phenomenon \citep{zuriguel2020contact}, and turbulence \citep{helbing2007dynamics}, to name a few.
A typical pedestrian conflict scenario involves cross flow. Interweaving of perpendicular pedestrian flows gives rise to the stripe phenomenon, a typical self-organizing pattern that helps reduce the "friction" of crossing flows. Further pedestrian experiments and the development of models on this topic were conducted \citep{cao2017fundamental,zanlungo2023macroscopica,mullick2022analysis,zanlungo2023macroscopicb}. Beyond these, research on individual pedestrian behavior has also been explored. Studies have revealed that crowds attempt to avoid longitudinal intrusions by employing lateral motions \citep{nicolas2019mechanical}. Pedestrians have been observed to cross through extremely dense environments successfully. These findings highlight the dynamics of pedestrian motion in dense crowds are hardly expressed by mechanical descriptions \citep{kleinmeier2020agent}. A field study concerning the "Love Parade Disaster" identified two types of pedestrians within dense crowds: active and inactive \citep{ma2015simulation}. Active pedestrians persistently create pathways in crowded conditions, while inactive individuals resort to localized movement driven by forces and spatial constraints, analogous to the findings presented by \citet{parisi2016experimental}.

In general, pedestrian crossing behaviors typically exhibit diverse characteristics, shaped by social norms. The most frequent motion within crowds is crossing behavior. For instance, in the subway, passengers often cross the crowd toward the exit, and during music festivals, audiences need to cross the crowd to reach the front, among other examples \citep{VideoExample}. Although these crossing phenomena are ubiquitous in daily life, but we currently possess limited knowledge. A potential reason lies in the significant risks associated with conducting controlled experiments involving dense crowds \citep{jin2019observational}. Therefore, to date, related research remains limited. To this end, a series of crowd-crossing experiments are established to conduct a comprehensive analysis of these frequently overlooked behaviors. The structure of this paper is organized as follows: In the next section, we will introduce the experimental setup. In Section\ref{section3}, the trajectory characteristics of crossing pedestrians will be analyzed. Subsequently, in Section\ref{section4}, the differential performances of pedestrian motion in static and dynamic contexts will be quantitatively presented, and the conclusions are provided in Section\ref{section5}.

\section{Experimental Setup} \label{section2}
\begin{figure}[ht!]
\nolinenumbers
\centering
\includegraphics[scale=0.16]{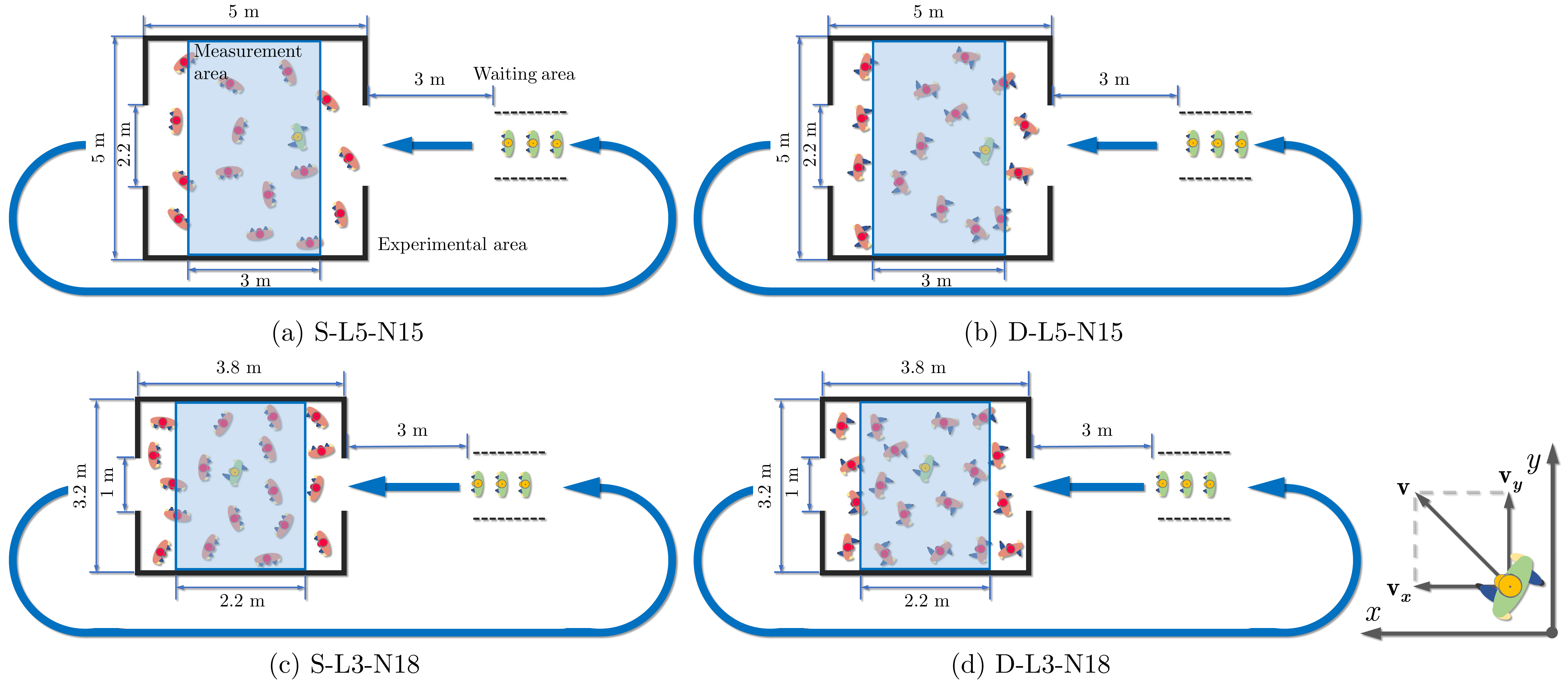}
\caption{Schematic illustration of experimental scenarios.}
\label{fig1}
\end{figure}

Experiments were conducted within the campus of Wuhan University of Technology, involving a total of 50 participants with a gender ratio of 1:1. The schematic diagram is presented in Fig.\ref{fig1}. In the experiment, participants sequentially crossed through the experimental area under two controlled conditions: global density and crowd contexts (static or dynamic). In the static context, the participants remained stationary within the experiment area during the crossing process, unable to move freely. Conversely, the dynamic context allowed participants to move freely within the experimental area. The experiment was conducted under both low-density and high-density conditions, with different configurations. The experimental area for the low-density (L5 class) was set at 5 m $\times$ 5 m, while the high-density (L3 class) was set at 3.8 m $\times$ 3.2 m. To ensure the mitigation of boundary effects, 3/5 of the central area was selected as the designated measurement area. Tab.\ref{table1} presents the experimental configuration, while Fig.\ref{fig1} illustrates the scenarios for both static and dynamic experiments. See \citet{wang2023} and the Data Availability section for details related to the experiments and data processing procedures.

\begin{table}[]
\nolinenumbers
\caption{Setup for crossing experiment.}
\centering
\begin{tabular}{*{6}{c}} 
    \toprule
    Index & Crowd context & Configuration (m\textsuperscript{2}) & 
    \makecell{Number of\\ Participants} & 
    \makecell{Global density\\ (ped/m\textsuperscript{2})} & 
    \makecell{Repetitions} \\
    \midrule
    S(D)-L5-N0  & Static (Dynamic) & 5$\times$5        & 0  & 0    & 30      \\
    S(D)-L5-N5  & Static (Dynamic) & 5$\times$5        & 5  & 0.2  & 50 (39) \\
    S(D)-L5-N10 & Static (Dynamic) & 5$\times$5        & 10 & 0.4  & 50 (39) \\
    S(D)-L5-N15 & Static (Dynamic) & 5$\times$5        & 15 & 0.6  & 49 (39) \\
    S(D)-L5-N20 & Static (Dynamic) & 5$\times$5        & 20 & 0.8  & 50 (40) \\
    S(D)-L5-N25 & Static (Dynamic) & 5$\times$5        & 25 & 1    & 49 (38) \\
    S(D)-L5-N30 & Static (Dynamic) & 5$\times$5        & 30 & 1.2  & 50 (39) \\
    S(D)-L5-N35 & Static (Dynamic) & 5$\times$5        & 35 & 1.4  & 49 (34) \\
    S(D)-L5-N40 & Static (Dynamic) & 5$\times$5        & 40 & 1.6  & 45 (35) \\
    S(D)-L5-N45 & Static (Dynamic) & 5$\times$5        & 45 & 1.8  & 51 (23) \\
    S(D)-L5-N49 & Static (Dynamic) & 5$\times$5        & 49 & 1.96 & 24 (20) \\
    S(D)-L3-N18 & Static (Dynamic) & 3.8$\times$3.2  & 18 & 1.48 & 29 (30) \\
    S(D)-L3-N24 & Static (Dynamic) & 3.8$\times$3.2  & 24 & 1.97 & 29 (29) \\
    S(D)-L3-N30 & Static (Dynamic) & 3.8$\times$3.2  & 30 & 2.47 & 29 (29) \\
    S(D)-L3-N36 & Static (Dynamic) & 3.8$\times$3.2  & 36 & 2.96 & 29 (29) \\
    S(D)-L3-N42 & Static (Dynamic) & 3.8$\times$3.2  & 42 & 3.45 & 16 (17) \\
    \bottomrule
\end{tabular}%
\label{table1}
\end{table}

\section{Trajectory Features} \label{section3}

\subsection{Trajectories} \label{subsection3.1}

\begin{figure}[ht!]
\nolinenumbers
\centering
\includegraphics[scale=0.33]{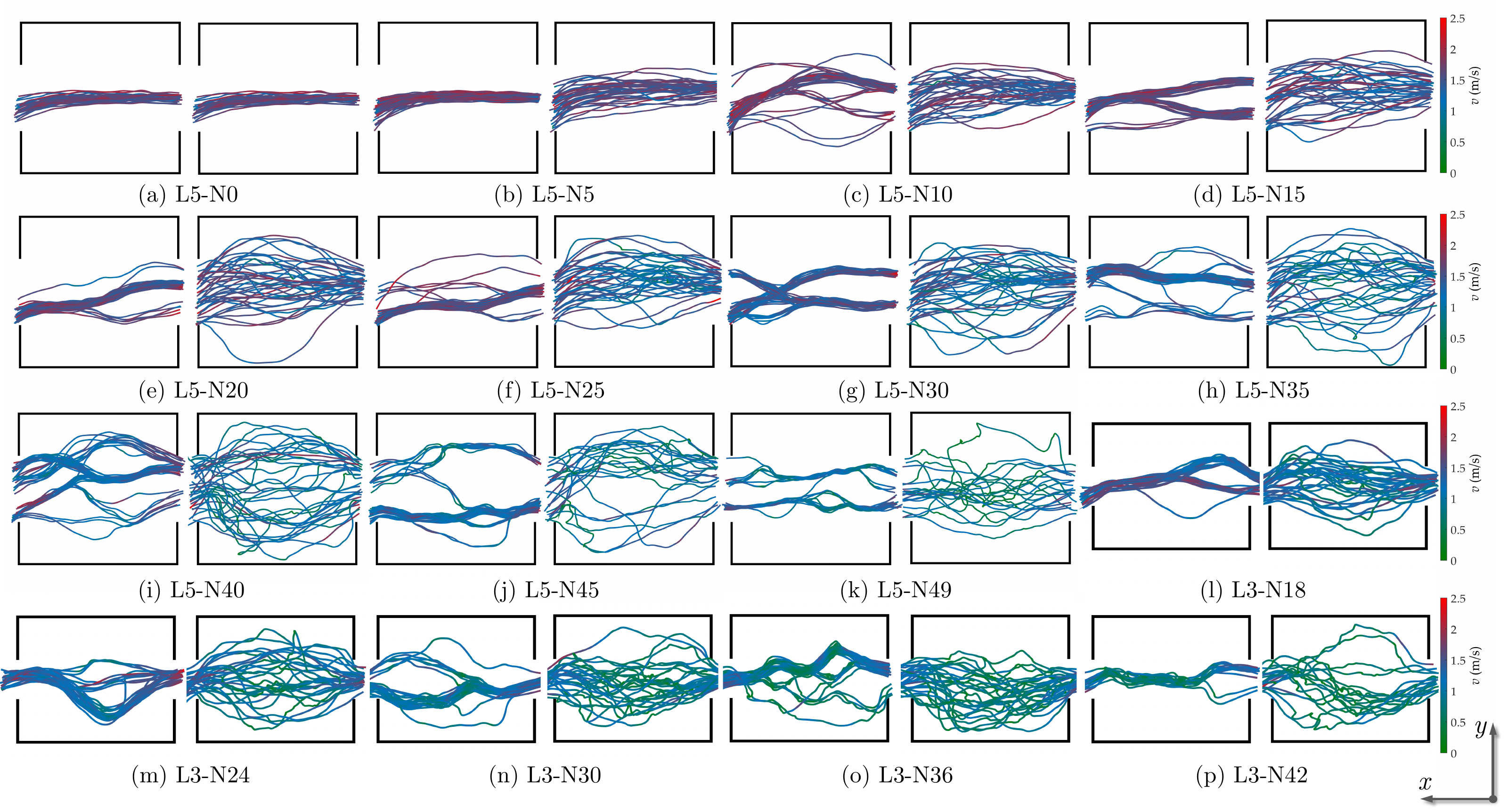}
\caption{ Pedestrian trajectory in static and dynamic crowds (on the left side of each subfigure: static experiment, on the right side: dynamic experiment. The positive direction of the \(x\)-axis in the lower right corner of the image indicates the pedestrian's direction of crossing).}
\label{fig2}
\end{figure}

\begin{figure}[ht!]
\nolinenumbers
\centering
\includegraphics[scale=0.78]{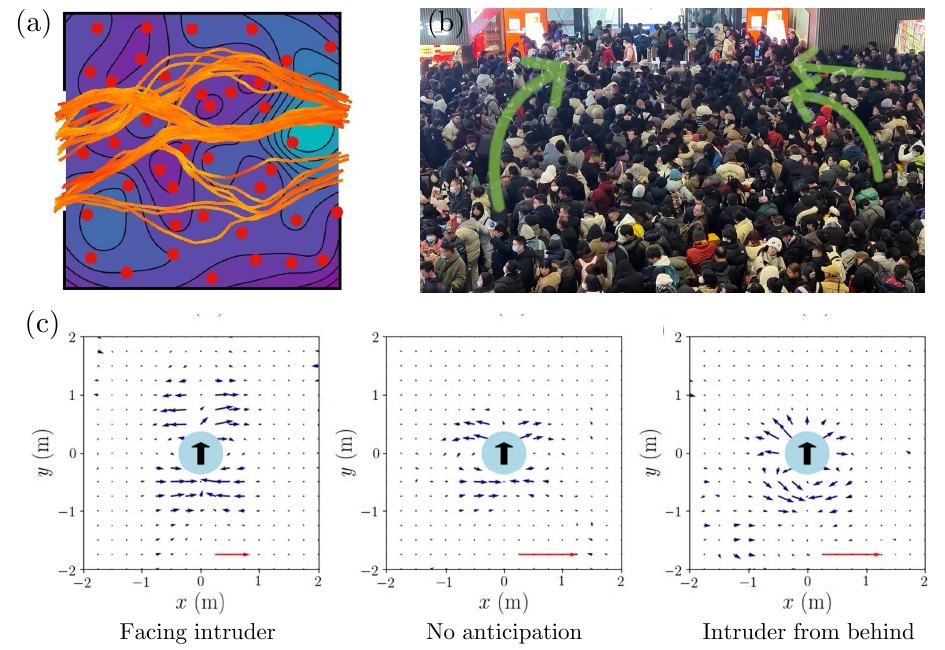}
\caption{(a) Cross-channel formation (experiment index: S-L5-N40), (b) Field observation of cross-channel formation in a train station, source: \citet{VideoExample}, (c) Local interactions of individual pedestrians crossing a static crowd, Source: \citet{nicolas2019mechanical}.}
\label{fig3}
\end{figure}

The trajectories of crossing pedestrians in each experiment are presented in Fig.\ref{fig2}, with the corresponding experiment index marked on each subfigure. In the static context, pedestrians crossed primarily via fixed channels, a phenomenon termed "cross-channel formation". In contrast, the trajectories appeared more stochastic in the dynamic context. The mechanisms of cross-interaction in static crowds have been investigated in experiments \citep{nicolas2019mechanical, kleinmeier2020agent} , simulation \citep{bonnemain2023pedestrians} and field observation \citep{VideoExample}, as shown in Fig.\ref{fig3}. This phenomenon demonstrates self-organization in crowd interactions, characterized by entropy reduction and a stable state. The mechanism of cross-channel formation varies depending on whether the crossing individuals are isolated or part of a continuous flow of pedestrians. Additionally, different interaction dynamics result in distinct phenomena. In the case of isolated individuals crossing a crowd, the static crowd dissipates in front of the pedestrian and converges behind them, resulting in a transient cross-channel that is only observed locally \citep{nicolas2019mechanical}. In comparison, for a continuous flow of pedestrians crossing, the cross channel tends to extend over a broader area and remain more stable, with dissipation typically taking longer \citep{VideoExample}, as illustrated in Fig.\ref{fig3}. Despite these differences, the underlying mechanism of cross-channel formation remains similar: improving crossing efficiency through spontaneous pedestrian ordering.

\subsection{Cross-channel Formation Mechanism} \label{subsection3.2}

\begin{figure}[ht!]
\nolinenumbers
\centering
\includegraphics[scale=0.25]{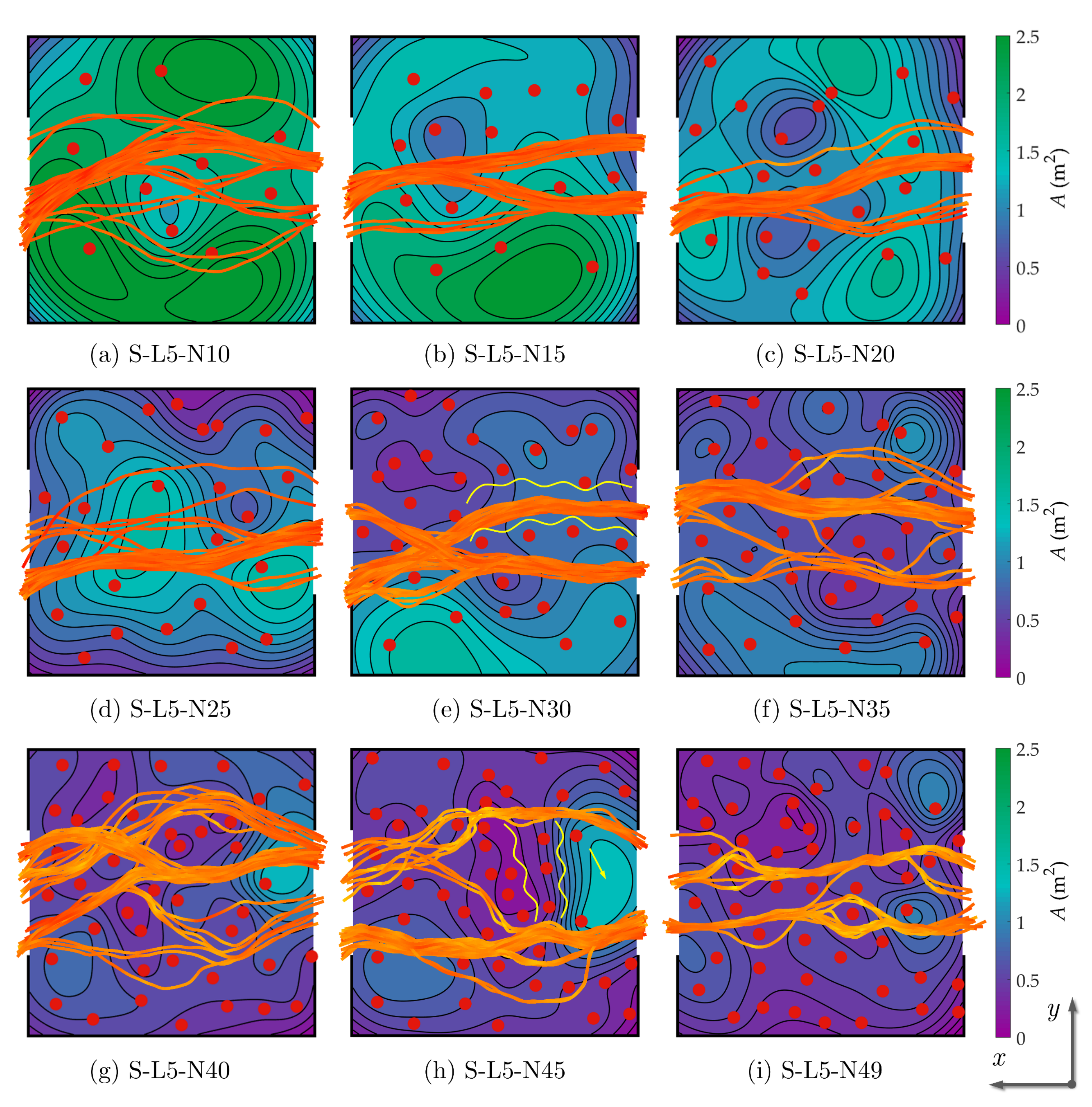}
\caption{The local area field diagrams in different experiments superimposed on the trajectories (the red dots in the diagram represent the distribution of static participants and the coordinate given in the lower right corner).}
\label{fig4}
\end{figure}

To explore the underground mechanism of cross-channel formation, we observed the evolution of channels within the local area field. The local area field, defined as the Thiessen polygon area occupied by individuals \citep{steffen2010methods}, was the intuitive metric of space during the pedestrian crossing process. Fig.\ref{fig4} illustrates the distribution of the local area field, spanning from experiment S-L5-N10 to experiment S-L5-N49. The orange curve represents the trajectory of each pedestrian. Subfigures (a)-(d) presented that pedestrians tend to select the shortest straight route when the density is low, resulting in trajectories without detours. Along with the global density increase, the local area field notably influences pedestrian trajectories. Pedestrians seem likely to opt for directions with higher local areas when detours are necessary. Despite the diverse mechanisms employed by individuals in selecting their routes, a predominant trend emerges wherein the majority of pedestrians converge upon some particular routes.

In high-density situations, pedestrians frequently make detours to plan their routes. Fig.\ref{fig4}(e) presents pedestrians crossing along orientations parallel to human walls rather than through them. Similarly, human walls will also obstruct the crossing route, forcing pedestrians to take wider detours. In the local area field depicted in Fig.\ref{fig4}(h), an empty area is present at the entrance (yellow arrow), making it a favorable choice for crossing. However, pedestrians plan their routes around the empty area due to the challenges posed by the two human walls behind it (yellow wavy line). These phenomena potentially illustrate that, in addition to transient interactions driven by spatial constraints (operational level), pedestrians tend to make local adjustments to achieve more optimal route decisions (tactical level).

\section{Comparisons\label{section4}}

The observation mentioned above has qualitatively analyzed the different characteristics of pedestrians crossing. In this section, we conducted a quantitative analysis of pedestrian behaviors and velocity
variation.

\subsection{Behavior Analysis\label{subsection4.1}}

To facilitate quantitative observation, we introduce the metric of the swarm factor as a range-based measurement method \citep{duives2015quantification, jia2022revisiting} to assess the local spatial dynamics. The swarm factor \((S)\) is defined to characterize the density of pedestrians in a pre-defined circular region, with the formula:

\begin{equation}\label{1}
{S} = \sum\limits_j {{n_j}(t)}
\end{equation}

\begin{equation}\label{2}
{n_j}(t) = \left\{ \begin{array}{l}
1\;\;{\rm{if}}\left\| {{\bm{x}_j}(t) - {\bm{x}_i}(t)} \right\| < r\\
0\;\;{\rm{otherwise}}
\end{array} \right.
\end{equation}

\begin{figure}[ht!]
\nolinenumbers
\centering
\includegraphics[scale=0.35]{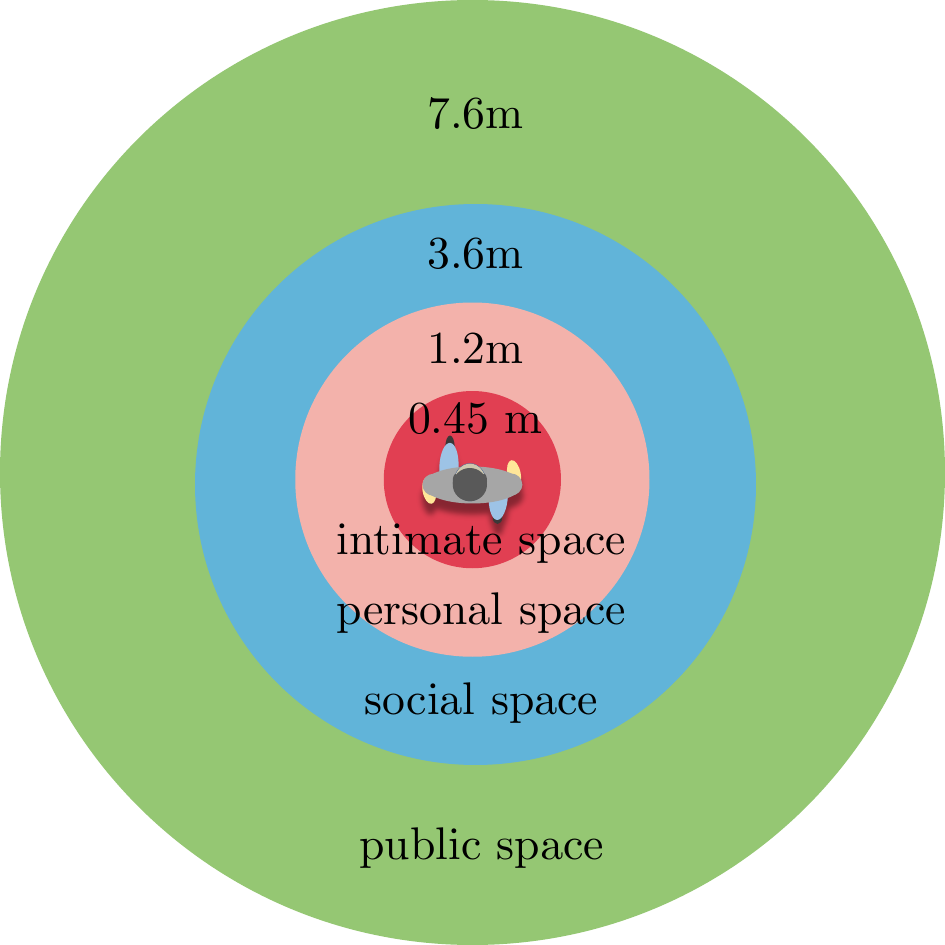}
\caption{Diagram of individual space requirement.}
\label{fig5}
\end{figure}

There, \(S\) represents the swarm factor of pedestrian \(i\) at moment \(t\), while \({{n}_{j}}(t)\) is a binary variable denotes for a given pedestrian \(i\), whether any pedestrian \(j\) enters the intimate space at moment \(t\). \({\bm{{x}}_{i}}(t)\) denotes the coordinate of pedestrian \(i\) at moment \(t\), similar conventions apply to \({\bm{{x}}_{j}}(t)\). \citet{hall1966hidden} defined four circular pedestrian space requirements, consisting of intimate distance, personal distance, social distance, and public distance, as shown in Fig.\ref{fig5}. During daily movement, people will avoid letting strangers enter their intimate space. Therefore, from a sociological perspective, the intimate distance can be considered as the critical distance during the crossing process. In this consideration, \(r\) is set to be 0.4 m, which is slightly smaller than the upper limit of intimate distance and usually considered as a critical value of pedestrian collision in modelling \citep{helbing1995social}. In this context, a non-zero swarm factor indicates that the pedestrian is in a state of "discomfort" concerning the setting of \(r\).

\begin{figure}[ht!]
\nolinenumbers
\centering
\includegraphics[scale=0.58]{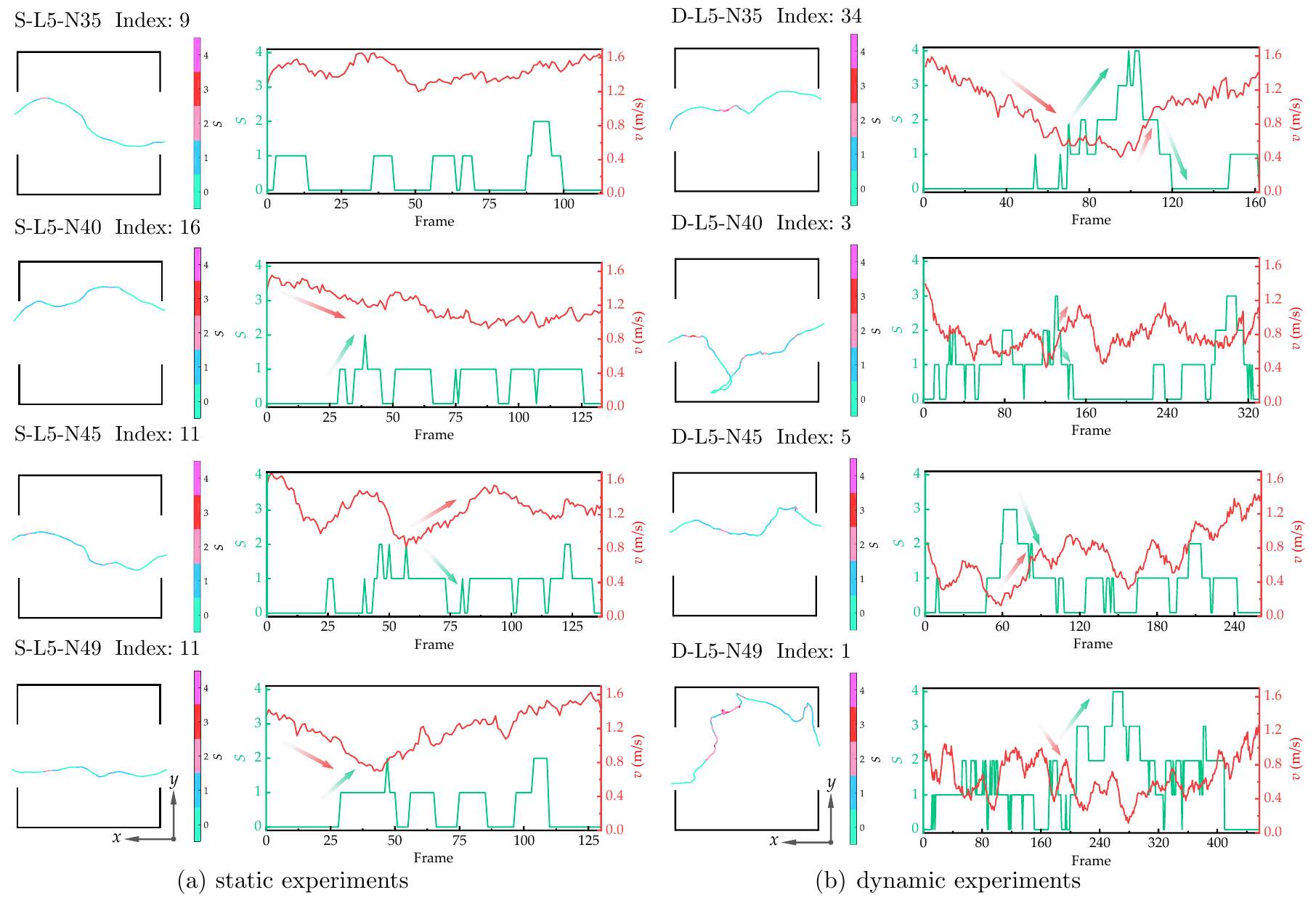}
\caption{Speed variation versus swarm factor of cross pedestrians in static and dynamic experiments, the subheadings indicate the pedestrian indexes, representing the trajectories of the respective experiments.}
\label{fig6}
\end{figure}

Based on the swarm factor introduced above, we conducted quantitative observations of pedestrian trajectories. Fig.\ref{fig6}(a) presents the variation of the swarm factor versus speed within the static context. Changes in pedestrian speed show a continuous increase and decrease rather than sudden transitions. Additionally, based on measurements of the swarm factor, the amplitude of the swarm factor and speed changes are limited during the crossing process. Comparing the swarm factor and speed variations reveals no distinctive features. We conclude that crossing pedestrians in a static crowd are less influenced by the environment than by crowd dynamics. Firstly, pedestrians actively avoid high-density areas during the crossing process. Moreover, the static environment allows them to plan their trajectories (inducing cross-channel formation). Consequently, pedestrians can maintain streamlined trajectories. In dynamic contexts, the temporal evolution of pedestrian speed versus the swarm factor is depicted in Fig.\ref{fig6}(b). In contrast to static contexts, both speed and swarm factor exhibit more pronounced fluctuations under dynamic conditions. These results demonstrate key differences in pedestrian crossing behavior: in static contexts, tactical behaviors are adopted to maintain stable motion when crossing, whereas in dynamic conditions, operational behaviors predominate, showing density-dependent behavioral adaptation (evidenced by inverse synchronization between speed and swarm factor).

\subsection{Velocity Analysis\label{subsection4.2}}

\subsubsection{Crossing Efficiency\label{subsubsection4.2.1}}
The average crossing speed (\(\bar v_x\)) denotes the average speed in the crossing direction and serves as the measure of crossing efficiency within the crowd, expressed as:

\begin{equation}\label{3}
{\bar v_x} = \frac{L}{{{t_{{\rm{out}}}} - {t_{{\rm{in}}}}}}.
\end{equation}

\begin{figure}[ht!]
\nolinenumbers
\centering
\includegraphics[scale=1]{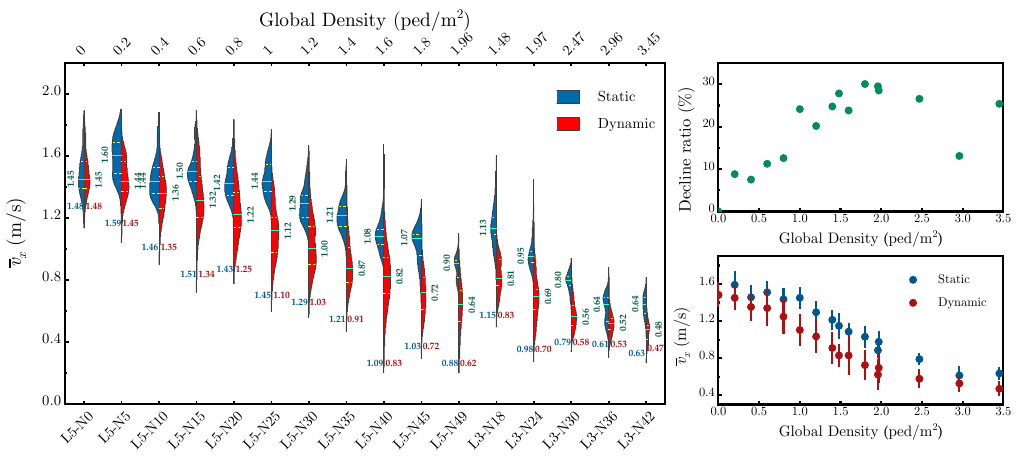}
\caption{Average crossing speed variation in different experimental groups, with median (thick green line and green numbers), upper and lower quartiles (yellow dotted line), and mean data (red and blue numbers). The subplots reflected the trend of the speed difference in static and dynamic contexts and the variation trend of mean and standard deviation($\text{mean} \pm \text{SD}$), respectively.}
\label{fig7}
\end{figure}

Here, \(L\) denotes the crossing length, \(t_{{\rm{in}}}\) indicates the moment of the pedestrian entering the experimental area, and \(t_{{\rm{out}}}\) represents the moment of the pedestrian leaving the experimental area. The distribution of average crossing speeds under static and dynamic contexts is presented in Fig.\ref{fig7}. The average crossing speed of the static crowd is significantly higher than the dynamic crowd. Within the static condition, the highest speed peak is observed in experiment S-L5-N5 (mean = 1.59 m/s), whereas the dynamic condition exhibits its speed peak in experiment D-L5-N0 (mean = 1.48 m/s). As global density increases (within 0-2 ped/m\textsuperscript{2}), the disparity in average crossing speed tends to widen, as shown in the subplots.

\subsubsection{Orthogonal Analysis\label{subsubsection4.2.2}}

\begin{figure}[ht!]
\nolinenumbers
\centering
\includegraphics[scale=0.85]{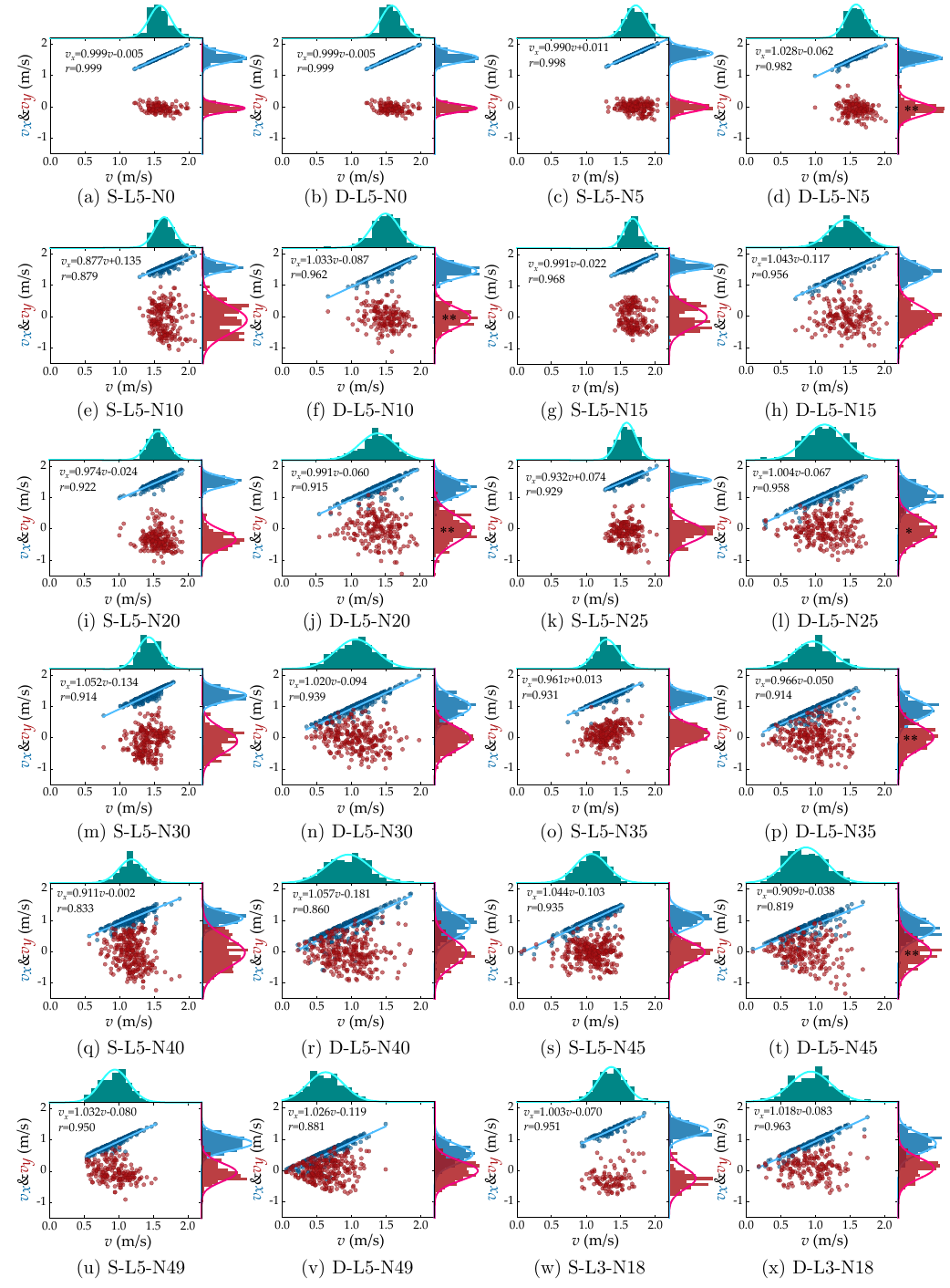}
\caption{Scatter plots and the corresponding marginal histograms of instantaneous velocities along orthogonal directions, *and ** indicate that, according to the Bartlett test, the transverse velocity data from static and dynamic contexts have significantly different variances at the 0.05 level and 0.01 level, respectively.}
\label{fig8}
\end{figure}

In this section, the instantaneous velocity is decomposed into two orthogonal components: the crossing direction (in the \(x\)-direction) and the transverse direction (in the \(y\)-direction). This decomposition allows us to investigate the variations in crossing velocity (\(v_x\)) and transverse velocity (\(v_y\)) so that different detour features within static and dynamic contexts can be quantitatively elucidated. The velocity deconstruction correspondence within each experiment is independently analyzed, as illustrated in Fig.\ref{fig8}. The results of Bartlett's test for homogeneity of variances in transverse velocity across experiments are presented in Tab.\ref{table2}. Fig.\ref{fig8} shows a consistent linear relationship between the crossing velocity (\(v_x\)) and the instantaneous velocity (\(v\)). Moreover, by comparing the distribution trends at different global densities, the transverse velocity (\(v_y\)) exhibits a systematic increase within the dynamic context (also see Tab.\ref{table2}), indicating that pedestrians have a higher frequency of detours.

\begin{figure}[ht!]\ContinuedFloat
\nolinenumbers
\centering
\includegraphics[scale=0.85]{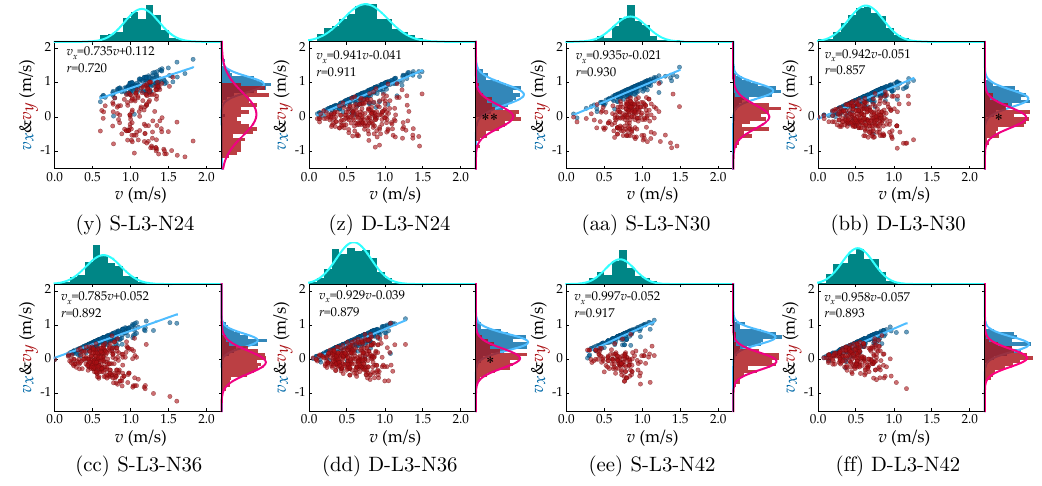}
\caption{(continued).
\label{fig8c}}
\end{figure}

\begin{table}[htbp]
\nolinenumbers
\centering
\caption{Results of Bartlett's test for homogeneity of variances, NS: not significant.}
\begin{tabular}{lrrlrrlrr}
\toprule
Index & $\chi^2$ & $p$-value & Index & $\chi^2$ & $p$-value & Index & $\chi^2$ & $p$-value \\
\midrule
S(D)-L5-N5  & 25.6737 & $p < 0.01$ & S(D)-L5-N30 &  0.5345 & NS & S(D)-L3-N18 &  0.1575 & NS \\
S(D)-L5-N10 & 22.1530 & $p < 0.01$ & S(D)-L5-N35 & 21.6510 & $p < 0.01$ & S(D)-L3-N24 & 70.1284 & $p < 0.01$ \\
S(D)-L5-N15 &  0.7035 & NS & S(D)-L5-N40 &  3.7621 & NS & S(D)-L3-N30 &  4.3785 & $p < 0.05$ \\
S(D)-L5-N20 & 30.1186 & $p < 0.01$ & S(D)-L5-N45 & 16.3658 & $p < 0.01$ & S(D)-L3-N36 &  6.4991 & $p < 0.05$ \\
S(D)-L5-N25 &  5.1891 & $p < 0.05$ & S(D)-L5-N49 &  1.0561 & NS & S(D)-L3-N42 &  0.0793 & NS \\
\bottomrule
\end{tabular}
\label{table2}
\end{table}

\begin{figure}[ht!]
\nolinenumbers
\centering
\includegraphics[scale=0.78]{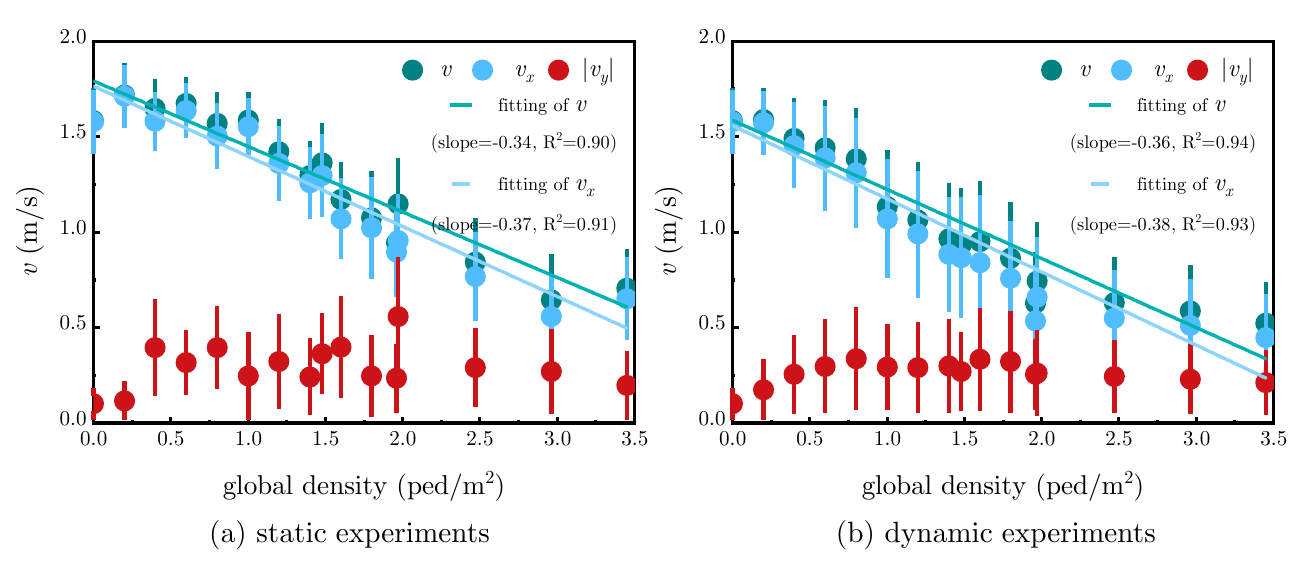}
\caption{Variation of instantaneous velocity (\(v\)), crossing velocity (\(v_x\)) and the absolute value of transverse velocity \(|v_y|\) versus global density. The dash and the shaded area represent the mean value and one-sigma error, respectively.}
\label{fig9}
\end{figure}

Fig.\ref{fig9} illustrates the corresponding variations in the instantaneous velocity (\(v\)), the crossing velocity (\(v_x\)), and the absolute value of transverse velocity (\(|v_y|\)). A comparison between Fig.\ref{fig9}(a) and Fig.\ref{fig9}(b) reveals that speed fluctuations expand with increasing global density. Moreover, in dynamic contexts, the decay trend of the crossing velocity (including instantaneous velocity) is more pronounced than in static contexts, as confirmed by statistical analysis. Most importantly, in both dynamic and static experiments, variations in speeds exhibit a fundamental pattern: as global density increases, both instantaneous velocity and crossing velocity decrease, while transverse velocity primarily remains within the interval of [0.2, 0.5]. This observation implies that even in highly congested scenarios, pedestrians may continue to strategically maintain motion through detours \citep{wang2023}.

\section{Conclusions} \label{section5}

This study investigated pedestrian crossing behavior within dynamic and static crowds. To explore discrepancies in pedestrian crossing behavior, a quantitative density metric termed the swarm factor was introduced. In static contexts, the limited variation in speed and swarm factor suggests that crossing pedestrians are less responsive to density increases, indicating potential tactical behavior that may contributes to the formation of cross-channels, a unique self-organization phenomenon. This contrasts with dynamic crowds, where speed and swarm factor synchronization inversely, indicating density-dependent behavioral adaptation (operational level). Furthermore, employing orthogonal analysis, we identified a fundamental pattern in pedestrian crossing: as global density increases, both instantaneous velocity and crossing velocity decrease, while transverse velocity remains stable. This finding suggests that detour behavior may be a potential mechanism for pedestrians to maintain dynamism within crowds. These observations contribute to the knowledge of pedestrian behavior patterns and provide an empirical understanding of crowd dynamics.

\section*{Data Availability}
The data can be found here: {\url{https://doi.org/10.34735/ped.2019.4}} (Pedestrian Dynamics Data Archive)
 or 
{\url{https://drive.google.com/drive/folders/1NYVnRp0z8VPuskfezMr51gB-sraOf6Iq?usp=drive_link}} (Google Drive).

\section*{Acknowledgments}
This work was supported by the National Natural Science Foundation of China (Grant No. 52072286, 71871189, 51604204), and the Fundamental Research Funds for the Central Universities (Grant No. 2022IVA108).

\bibliographystyle{aasjournal}

\end{document}